\documentclass[aps,prl,twocolumn,superscriptaddress,nofootinbib]{revtex4-1}
\usepackage{graphicx}
\usepackage{amsmath}
\usepackage{subfigure}
\usepackage{braket}
\usepackage{color}
\begin{document}

%\preprint{}

\title{Linear probing of molecules at micrometric distances from a surface with sub-Doppler frequency resolution}


\author{J. Lukusa Mudiayi}
\affiliation{Laboratoire de Physique des Lasers, Universit{\'e} Sorbonne Paris Nord, F-93430, Villetaneuse, France}
\affiliation{CNRS, UMR 7538, LPL, 99 Avenue J.-B. Cl{\'e}ment, F-93430 Villetaneuse, France}

\author{I. Maurin }
%\author{}
\affiliation{Laboratoire de Physique des Lasers, Universit{\'e} Sorbonne Paris Nord, F-93430, Villetaneuse, France}
\affiliation{CNRS, UMR 7538, LPL, 99 Avenue J.-B. Cl{\'e}ment, F-93430 Villetaneuse, France}

\author{T. Mashimo}
\affiliation{Laboratoire de Physique des Lasers, Universit{\'e} Sorbonne Paris Nord, F-93430, Villetaneuse, France}
\affiliation{CNRS, UMR 7538, LPL, 99 Avenue J.-B. Cl{\'e}ment, F-93430 Villetaneuse, France}
%\affiliation{Department of Physics, Chuo University, Kasuga, Bunkyo-ku, Tokyo 112-8551, Japan}

\author{J. C. de Aquino Carvalho}
\affiliation{CNRS, UMR 7538, LPL, 99 Avenue J.-B. Cl{\'e}ment, F-93430 Villetaneuse, France}
\affiliation{Laboratoire de Physique des Lasers, Universit{\'e} Sorbonne Paris Nord, F-93430, Villetaneuse, France}
%\affiliation{Department of Physics, Chuo University, Kasuga, Bunkyo-ku, Tokyo 112-8551, Japan}

\author{D. Bloch }
%\author{}
\affiliation{CNRS, UMR 7538, LPL, 99 Avenue J.-B. Cl{\'e}ment, F-93430 Villetaneuse, France}
\affiliation{Laboratoire de Physique des Lasers, Universit{\'e} Sorbonne Paris Nord, F-93430, Villetaneuse, France}

\author{S. K. Tokunaga }
%\author{}
\affiliation{Laboratoire de Physique des Lasers, Universit{\'e} Sorbonne Paris Nord, F-93430, Villetaneuse, France}
\affiliation{CNRS, UMR 7538, LPL, 99 Avenue J.-B. Cl{\'e}ment, F-93430 Villetaneuse, France}

\author{B. Darqui\'e }
%\author{}
\affiliation{CNRS, UMR 7538, LPL, 99 Avenue J.-B. Cl{\'e}ment, F-93430 Villetaneuse, France}
\affiliation{Laboratoire de Physique des Lasers, Universit{\'e} Sorbonne Paris Nord, F-93430, Villetaneuse, France}

\author{A. Laliotis }
%\author{}
\email{laliotis@univ-paris13.fr}
\affiliation{Laboratoire de Physique des Lasers, Universit{\'e} Sorbonne Paris Nord, F-93430, Villetaneuse, France}
\affiliation{CNRS, UMR 7538, LPL, 99 Avenue J.-B. Cl{\'e}ment, F-93430 Villetaneuse, France}

%\thanks{}


\date{\today}

\begin{abstract}
 
We report on precision spectroscopy of sub-wavelength confined molecular gases. This was obtained by rovibrational selective reflection of $\mathrm{NH_3}$ and $\mathrm{SF_6}$ gases using a quantum cascade laser at $\mathrm{\lambda \approx 10.6}$ $\mathrm{\mu m}$. Our technique probes molecules at micrometric distances ($\mathrm{\approx\lambda/2\pi}$) from the window of a macroscopic cell with sub-MHz resolution, allowing molecule-surface interaction spectroscopy. We exploit the linearity and high-resolution of our technique to gain novel spectroscopic information on the $\mathrm{SF_6}$ greenhouse gas, useful for enriching molecular databases. The natural extension of our work to thin-cells will allow compact frequency references and improved measurements of the Casimir-Polder interaction with molecules.

\end{abstract}

%\keywords{out of equilibrium}

\maketitle
%\section{}
%\subsection{}
%\subsubsection{}
%\section{Introduction}

High-resolution molecular spectroscopy in gas cells has far-reaching applications ranging from Earth and atmospheric sciences~\cite{Polyansky_PRL_2015,vaskuri_uncertainty_2018} to astrophysics~\cite{roueff_spectroscopic_2020}, metrology and frequency referencing \cite{gilbertspie2001, Benabid_nature_2005,zektzer_chipscale_2020,santagata_high-precision_2019}, gas sensing and trace detection~\cite{galli_spectroscopic_2016}, as well as fundamental physics measurements~\cite{Daussy_PRL_2008, Moretti_PRL_2013, Daussy_PRL_1999, diouf_lamb-peak_2020, Cournol_2019}. The growing demand for miniaturisation has led to the fabrication of compact platforms that interface molecular gases with solid-state devices, such as on-chip waveguides \cite{zektzer_chipscale_2020}, hollow core fibres \cite{Benabid_nature_2005}, porous media \cite{Svensson_PRL_2011} and thin cells  \cite{Hartmann_PRA_2016}. However, the above experiments typically operate at high gas pressures due to low transition probabilities of molecular lines and the resolution is limited either by pressure or Doppler broadening. Therefore, achieving precision spectroscopy of a confined molecular gas is challenging. 

Confined gases have been studied primarily with atomic alkali vapors with implications ranging from fundamental physics to quantum technologies. Thin cells have been used to study the Dicke narrowing effect \cite{romer1955}, allowing high-resolution sub-Doppler linear spectroscopy \cite{Dutier_EPL_2003}, and more recently, investigations of dipole-dipole interactions with high density atomic vapors~\cite{Peyrot_PRL_2018}. Probing Rydberg atoms in thin cells is also a promising approach for quantum information processing \cite{Ripka446}. Three dimensional atomic confinement has also been explored with photonic crystals~\cite{Ballin_APL_2013} or random media~\cite{Villalba_OptLett_2013}. Finally, the fundamental Casimir-Polder interaction has been studied with confined atomic vapors, either in nanometric thin cells \cite{fichet_epl_2007}, or by selective reflection spectroscopy~\cite{Bloch_Review_2003}.

High-resolution spectroscopy of confined molecules offers attractive prospects for fabricating compact frequency references. Additionally, it paves the way for spectroscopic probing of the Casimir-Polder molecule-surface interaction, a topic of interest for physical-chemistry or atmospheric sciences \cite{Fiedler_PCCP_2019, AntezzaPRB2020}, as well as for fundamental physics due to the rich geometry of polyatomic molecules. The dependence of the Casimir-Polder interaction on molecular orientation (anisotropy) \cite{Thiyam_PRA_2015, Bimonte_PRA_2016, AntezzaPRB2020} or on molecular chirality  \cite{Butcher_NJP_2012} (when the surface is also chiral), are for instance open theoretical questions. Although molecule-surface interactions are of fundamental interest, so far experimental tests are few and comparison with theoretical predictions has been challenging \cite{Raskin1974, Raskin1969, Boustimi_2001, brand_2015, wagner_natcommun_2014}.

One possible way for probing molecular gases close to a dielectric surface, in an effectively confined environment, is via selective reflection in a molecular gas cell \cite{Woerdman_1975,Schuurmans1975}. Frequency modulated selective reflection spectroscopy (FMSR), under normal incidence, is sensitive to particles (molecules or atoms) that move parallel to the dielectric window of the gas cell at distances comparable to the reduced wavelength of excitation ($\mathrm{\lambdabar}=\lambda/2\pi$). This feature has made FMSR an important spectroscopic technique for measuring Casimir-Polder interactions with excited states atoms \cite{ducloy_jphysique_1991, chevrollier1992}. Additionally, FMSR is a linear spectroscopy without cross-over resonances that often muddle saturated absorption spectra. This allows easy interpretation of observed lineshapes and the study of gas properties, such as collisional shifts and broadenings \cite{akulshin_1982, vuletic_optcomm_1993, laliotisAPB2008, chevrollier1992}, even at high densities where volume absorption spectroscopy is unfeasible. The above advantages of FMSR can have an important impact for molecular spectroscopy \cite{Polyansky_PRL_2015,vaskuri_uncertainty_2018,roueff_spectroscopic_2020,galli_spectroscopic_2016, Faye_JMS_2018, Khalil1999,krieg2005} allowing simultaneous measurements of transition frequencies, intensities and collisional broadenings. Although attempts have been made to probe alkali dimers \cite{SHMAVONYAN201514} at high temperatures, high resolution FMSR has so far been exclusively performed on atomic vapors.  

Here, we perform high resolution rovibrational FMSR of $\mathrm{NH_3}$ and $\mathrm{SF_6}$ gases at $\mathrm{\lambda \approx 10.6\;\mu m}$. Our experiment probes molecules at a depth of about $\mathrm{\lambdabar \approx 1.7\;\mu m}$ with a sub-MHz resolution, limited by the linewidth of our quantum cascade laser (QCL) source. The exceptional combination of linearity and high-resolution offered by selective reflection is used to resolve the hyperfine structure of $\mathrm{NH_3}$ and gain novel spectroscopic information on the $\mathrm{SF_6}$ molecule, of importance to atmospheric physics and corresponding molecular databases. Finally, we use FMSR to perform molecule-surface interaction spectroscopy with a sensitivity in Casimir-Polder shifts of about 10 kHz at 1 $\mathrm{\mu m}$ from the surface.   

Selective reflection is performed on a gas cell at room temperature constructed out of metallic vacuum tubes with ZnSe windows. We use a commercial QCL, of output power $\approx$ 5 mW, whose frequency is scanned by changing the laser current. A frequency modulation (FM) is applied at $\mathrm{f_{FM}\approx8}$ kHz with a peak-to-peak amplitude $\mathrm{M\approx0.5}$ MHz. The reflection from the gas/window interface (Fig.1) is demodulated at frequency $\mathrm{f_{FM}}$ with a lock-in amplifier. In the limit $\mathrm{M \ll \Gamma}$ where $\mathrm{\Gamma}$ is the homogeneous linewidth, this provides the derivative of the direct signal and increases the contrast of the sub-Doppler contribution originating from molecules that are slow in the direction of the beam \cite{ducloy_jphysique_1991}. To avoid any residual Doppler broadening, selective reflection is performed at normal incidence. When  $\mathrm{\Gamma \ll \Delta_{D}}$, where $\Delta_{D}$ is the Doppler full-width-at-half-maximum (FWHM), the FMSR signal ($\mathrm{S_{FMSR}}$) becomes a dispersive Lorentzian curve of width equal to $\mathrm{\Gamma}$ \cite{akulshin_1982}:

\begin{equation}
\mathrm{S_{FMSR} \propto -\mu^2 M \frac{N}{\Gamma} \frac{\lambdabar}{u_{p}}\frac{\frac{2\delta}{\Gamma}}{1+\left( \frac{2\delta}{\Gamma} \right)^2}}
\label{Eq1}
\end{equation}
Here, $\mathrm{S_{FMSR}}$ is defined as the FM demodulated signal normalized by the off-resonant reflection; $\mu$ is the dipole moment matrix element of the probed transition, N is the population of the lower state, $\mathrm{u_{p}}$ is the most probable molecular velocity in the direction of the beam and $\mathrm{\delta}$ is the laser detuning.  

We focus on $\mathrm{NH_3}$ and $\mathrm{SF_6}$ molecules with strong rovibrational transitions in the mid-infrared ($\mathrm{\approx 10.6\; \mu m}$) window. Ammonia ($\mathrm{NH_3}$) is of tetrahedral geometry with widely spaced rotational levels. The only major line of the most abundant ammonia isotope within the 150 GHz spectral window of our laser is the saP(1,0) rovibrational transition at $\mathrm{28,427,281.4}$ MHz~\cite{HITRAN2016}, from the ground state to the first $\mathrm{\nu_2}$ vibration. The spread of the observed hyperfine structure, resulting from electric quadrupole interactions in the lower level, is a few MHz \cite{URBAN_2000} (Fig.2) and is unresolved with Doppler limited resolution ($\mathrm{\Delta_{D}=85\; MHz}$ at room temperature). Sulfur hexafluoride ($\mathrm{SF_6}$) is of spherical geometry, presenting a multitude of transitions of the $\mathrm{\nu_3}$ vibrational mode in the frequency range of our laser with a superfine structure occasionally resolved even with Doppler limited resolution ($\mathrm{\Delta_{D}=29\; MHz}$ at room temperature)~\cite{Bobin_JMS_1987, acef_new_2000}.

A saturated absorption set-up provides a molecular frequency reference in the volume at low gas pressure, allowing a frequency calibration of our scans. Saturated absorption is recorded simultaneously with selective reflection and is demodulated with a lock-in amplifier at $\mathrm{f_{FM}}$ or $\mathrm{2f_{FM}}$. $\mathrm{2f_{FM}}$ demodulation provides a better contrast of the narrow peaks at the expense of signal amplitude. The frequency drift of the free running QCL is incompatible with high resolution spectroscopy. We thus use an auxiliary set-up to lock the laser frequency either on the side of a direct absorption profile, or at the slope of the first derivative of the linear absorption of the ammonia saP(1,0) transition. The laser frequency is then scanned by adding an offset to the error signal. The laser stabilization circuit only corrects the slow laser frequency drift (timescale $\mathrm{> 1ms}$) and not the laser linewidth.     

\begin{figure}[!t]
\includegraphics[width=85mm]{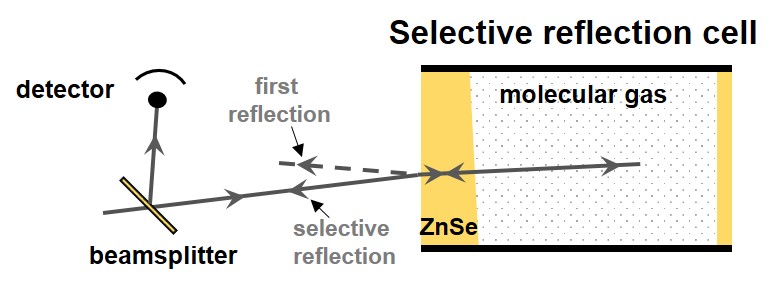}%
\caption{Schematic of the selective reflection measurement. The wedged window allows separation of the selective reflection beam (solid line at normal incidence) from the first reflection (dashed line).
\label{Fig2}}
\end{figure}

\begin{figure}[!t]
\includegraphics[width=85mm]{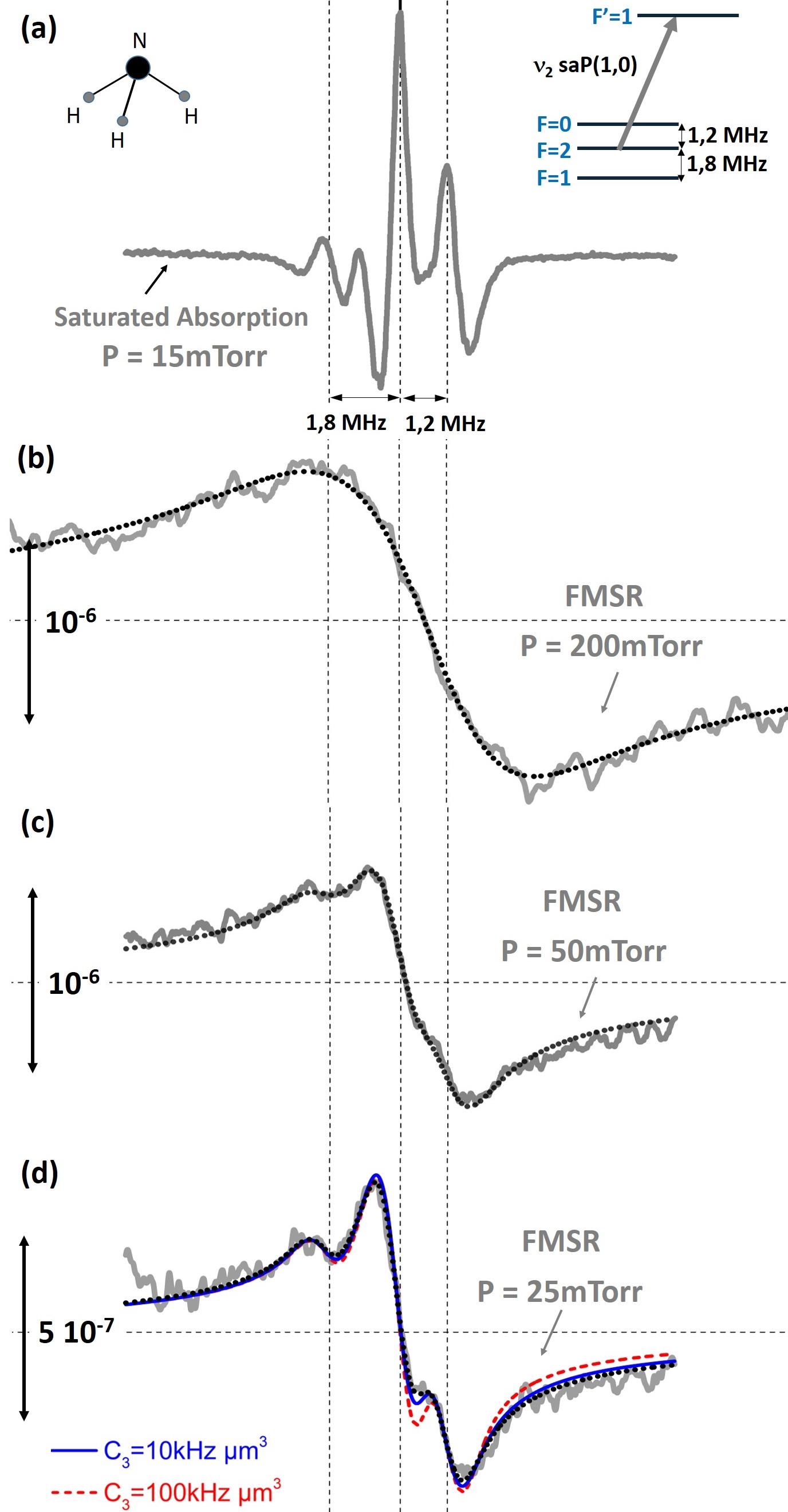}%
\caption{Selective reflection of ammonia $\nu_2$ saP(1,0) rovibrational line. (a) Saturated absorption reference (second harmonic detection). (b),(c),(d) Demodulated FMSR signal normalized by off-resonant reflectivity at P=200 mTorr, P=50 mTorr, P=25 mTorr respectively. The FM amplitude is M$\approx$0.5 MHz. Black dotted lines are fits of a theoretical model including the effects of frequency modulation and laser linewidth. In (d) we also show fits that include molecule-surface interactions with $\mathrm{C_3=10}$ kHz $\mathrm{\mu m^3}$ (solid blue line) and $\mathrm{C_3=100}$ kHz $\mathrm{\mu m^3}$ (dashed red line). 
Top right: level structure of the $\nu_2$ saP(1,0) rovibration. The position of hyperfine transitions are indicated as dashed vertical lines.
\label{Fig2}}
\end{figure}

Fig.2 shows selective reflection spectra of $\mathrm{NH_3}$ at various gas pressures (from P=25 mTorr to P=200 mTorr) along with a saturated absorption reference (Fig.2a). The hyperfine structure of the lower rovibrational level and crossover resonances are visible on the saturated absorption spectrum, typically recorded at $\mathrm{P=15\;mTorr}$. The hyperfine structure of ammonia is also resolved by FMSR for pressures lower than $\approx$50 mTorr (Fig.2c,d). However, unlike for saturated absorption spectroscopy, there are no crossover resonances and the ratio between the amplitudes of each $\mathrm{F\rightarrow F^{\prime}}$ hyperfine transition is defined by its theoretical estimated strength (1:5:3 for $\mathrm{0\rightarrow1,2\rightarrow1,1\rightarrow1}$ respectively \cite{URBAN_2000}). The FMSR signals of Fig.2 are the result of averaging 40 individual $\approx$2 min long scans. 

The frequency resolution of both saturated absorption and FMSR is determined by the laser linewidth, pressure broadening and FM excursion, whereas power and transit-time broadening have a minor effect in these conditions. The laser linewidth is $\mathrm{\approx0.6\;MHz}$ FWHM, experimentally measured by examining the saturated absorption linewidth while reducing FM distortions and pressure broadening \cite{Junior}. At low molecular pressures, laser linewidth is the frequency resolution limit for both techniques in the current set-up. 

At sufficiently high pressures, the FMSR signal linewidth is dominated by collisional broadening which is proportional to gas pressure. In this case the FMSR amplitude (about $10^{-6}$) remains constant with pressure because the reduction of lower state population is compensated by the decrease of transition linewidth (the ratio $\mathrm{N/\Gamma}$ in eqn.1 stays constant). This is seen in Fig.2a,b and was also verified for ammonia pressures as high as a few Torr \cite{Junior}. The loss of FMSR signal amplitude at lower pressures is a consequence of the laser linewidth limited frequency resolution (Fig.2d). In the experimental conditions of Fig.2 the  2$f_{FM}$ saturated absorption amplitude is about 1 order of magnitude larger than FMSR, however, the pressure range of saturated absorption is limited mainly because the saturation intensity increases with transition linewidth.

The black dotted curves of Fig.2 show the predicted FMSR lineshapes, including the exact FM lineshape distortion \cite{ducloy_jphysique_1991} (beyond the assumptions of eqn.1) and the effects of laser linewidth, considered to be a Gaussian function of $\mathrm{\approx0.6\;MHz}$ FWHM. The curves are adjusted for an overall amplitude (the ratio of the hyperfine components is fixed to its theoretical value), a collisional linewidth and shift (compared to the saturated absorption reference) and an offset. The deduced pressure broadening is about 27 MHz/Torr (FWHM) consistent with other values reported in literature \cite{Mejri_2015}, while the shift between FMSR and saturated absorption remains negligeable. 

We also use FMSR in order to measure molecule-surface, Casimir-Polder interactions. In Fig.2d we show the theoretical spectra including molecule-surface interaction effects \cite{ducloy_jphysique_1991}, adjusted for an overall amplitude and offset (blue and dashed red curves). We assume a $\mathrm{-C_3/z^3}$ potential, where z is the molecule-surface distance and $\mathrm{C_3}$ is the spectroscopic van der Waals coefficient (the difference between $\mathrm{C_3}$ coefficients of the probed states). Our FMSR spectroscopic Casimir-Polder measurement gives an upper bound of $\mathrm{\approx 10\;kHz\; \mu m^3}$ to the $\mathrm{C_3}$ coefficient. Systematic errors are reduced by the elimination of a parasitic background and laser frequency drift. There are no theoretical calculations for the spectroscopic $\mathrm{C_3}$ of ammonia that depends on the allowed electronic and vibrational contributions \cite{Buhmann_PRA_2012, brand_2015, wagner_natcommun_2014} and on the anisotropy due to the molecular rotation \cite{Bimonte_PRA_2016}. Nevertheless, based on previous calculations for other molecules \cite{Buhmann_PRA_2012, brand_2015, Bimonte_PRA_2016} we estimate that $\mathrm{C_3}$ should be less than $\mathrm{1\;kHz \; \mu m^3}$. We note that Casimir-Polder retardation can also play a role in such experiments \cite{Buhmann_PRA_2012}. The sensitivity of dedicated molecule-surface FMSR spectroscopy can be improved by reducing the QCL linewidth \cite{Cappelli_OL, Sow_APL_2014, Argence_NatPhot_2015} or by probing smaller wavelength transitions. For more details see the supplementary materials.

\begin{figure}[!t]
\includegraphics[width=85mm]{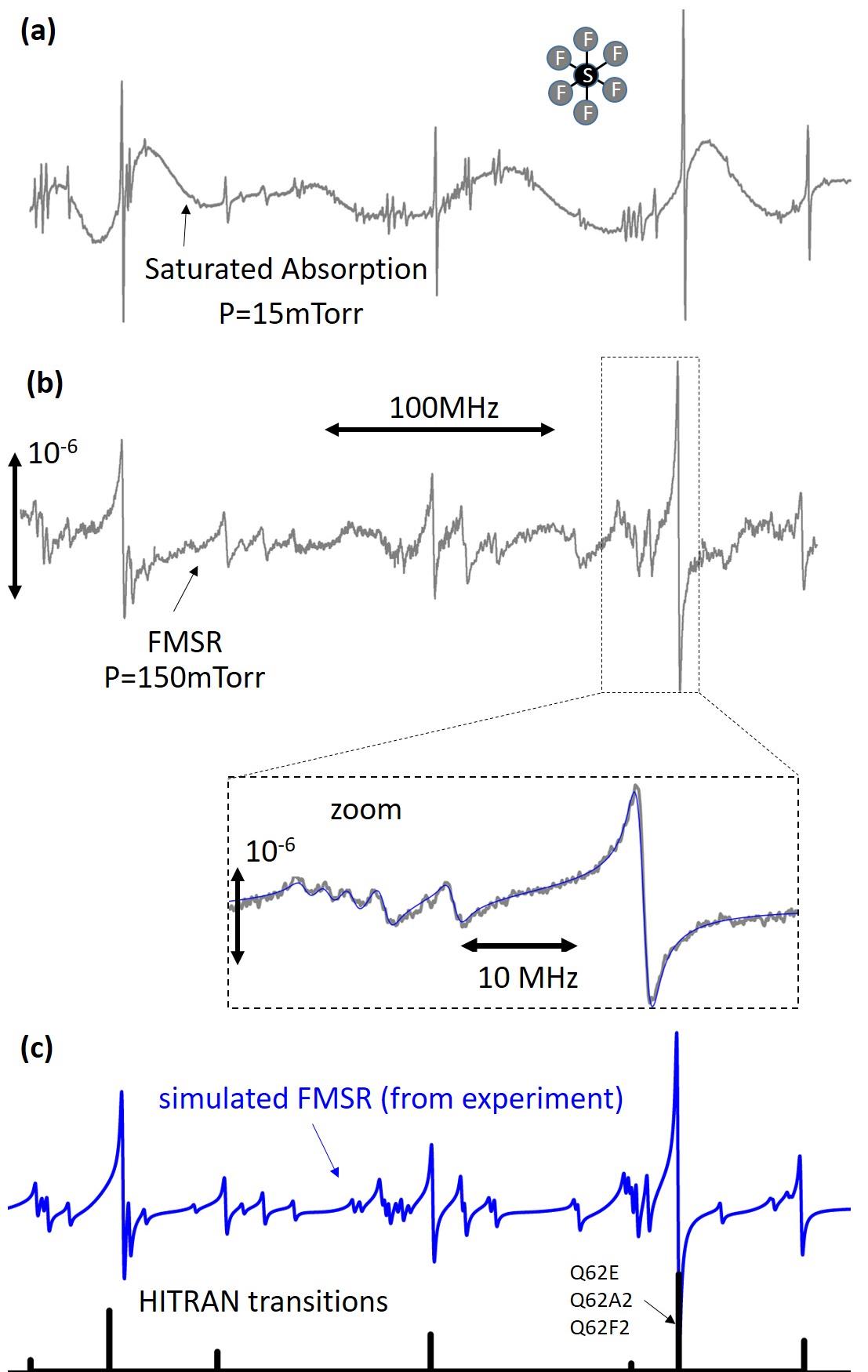}
\caption{Selective reflection of sulfur hexafluoride. (a) Saturated absorption (first harmonic detection) at P=15 mTorr.(b) Demodulated FMSR signal normalized by off-resonant reflectivity at 150 mTorr pressure with FM amplitude M=0.5 MHz. The spectrum is a patch of curves from 8 different regions each covering about 30-40 MHz. The scan of each region results from averaging 40 individual $\approx $2 min scans. A zoom of the FMSR scan is also shown around the Q62E, Q62A2 and Q62F2 $\mathrm{SF_6}$ transitions of the $\nu_3$ vibrational mode, predicted to be degenerate at 28,427,502.6 MHz in the HITRAN database~\cite{HITRAN2016}. A fit of the zoomed selective reflection spectra is shown as a blue solid line. (c) Predicted selective reflection spectra using the $\mathrm{SF_6}$ transition positions and amplitudes extracted from the FMSR experiment (solid blue line). The positions and amplitudes of the HITRAN listed rovibrations \cite{HITRAN2016} are also shown as vertical black lines. 
\label{Fig6}}
\end{figure}

We subsequently expanded our studies to $\mathrm{SF_6}$ for which molecular databases are incomplete \cite{Faye_JMS_2018}, because its dense rovibrational spectrum is difficult to resolve with traditional Fourier transform spectroscopy. High resolution saturated absorbtion measurements of $\mathrm{SF_6}$ have also been performed with $\mathrm{CO_2}$ lasers. However, these measurements were strongly limited to the parts of the 10.6 $\mathrm{\mu m}$ spectrum \cite{Bobin_JMS_1987, acef_new_2000} that are accessible with $\mathrm{CO_2}$ sources. Here, we perform selective reflection spectroscopy on $\mathrm{SF_6}$ rovibrations in the previously unexplored frequency range centered around the saP(0,1) transition of ammonia.

Fig.3 shows our experimental selective reflection results at a pressure of 150 mTorr, along with the saturated absorption reference taken at $\approx$ 15 mTorr. The HITRAN database is incomplete for $\mathrm{SF_6}$ transitions \cite{HITRAN2016}. We therefore used a Bristol Instruments wavemeter with a relative frequency uncertainty of $\approx$5 MHz to pinpoint the frequency positions of the $\mathrm{SF_6}$ rovibrations. Long frequency scans, such as the $\approx$ 300 MHz scan of Fig.3 can suffer from an oscillating background due to interference of the selective reflection signal with other parasitic reflections originating from various parts of the set-up. In order to reduce this background, we have used a system of electronic valves allowing us to empty and refill the chamber with molecules within tens of seconds and detect the difference between the two signals. Using this technique the interferometric background is reduced to values below $2.10^{-7}$ (see Supplementary materials).
    
The saturated absorption spectrum  provides a higher signal to noise ratio, however, the transition amplitudes cannot be easily extracted due to non-linearity, differential saturation between lines and the possible existence of cross-over resonances. Conversely, the amplitudes of the resolved molecular transitions can be extracted using linear selective reflection. For this purpose, we perform many local fits (see zoom in Fig.3a), delimiting a small part of the selective reflection spectrum, to minimize the effects of the interferometric background. The fits provide the relative amplitude, frequency position and pressure broadened linewidth of the transitions. Here, the effects of molecule-surface interactions are ignored. The uncertainty on the relative transition amplitudes depends on the short term noise of our experiment, which is smaller than $\mathrm{10^{-7}}$ (Fig.3b). This translates to $\approx 10\%$ uncertainty for the strongest transitions (see Supplementary Materials). Further improvements can be made by increasing the scan integration time. Within the resolution of our experiment (limited by laser linewidth), no collisional frequency shift is measurable between the positions obtained by FMSR and saturated absorption. The final simulated spectrum is shown in Fig.3c (solid blue curve). For comparison we also show the positions and amplitudes of $\mathrm{SF_6}$ transitions listed in the HITRAN database \cite{HITRAN2016} (black bars). The HITRAN data are insufficient to interpret our experimental curves. We can gain further information by expanding these measurements in the entire spectrum of the $\nu_{3}$ rovibration of $\mathrm{SF_6}$.     

Accurate determination of transition amplitudes and positions in laboratory experiments is crucial in atmospheric physics applications, in particular for monitoring gas concentrations by remote sensing experiments \cite{Polyansky_PRL_2015, vaskuri_uncertainty_2018}. Measuring the atmospheric abundance of $\mathrm{SF_6}$, an important greenhouse gas, is essential for monitoring global warming \cite{Faye_JMS_2018, Khalil1999}. However, $\mathrm{SF_6}$  and other heavy atmospheric species ($\mathrm{ClONO_2}$, $\mathrm{CF_4}$,...) with low-lying vibrational modes exhibit both a dense rotational structure and many hot bands, and traditional Fourier transform infrared spectra do not show isolated lines, but rather unresolved clusters of many transitions. For such species, molecular databases neglect many significant hot bands, feature inaccurate intensities, and cannot be exploited for atmospheric quantification \cite{Faye_JMS_2018, krieg2005}. In this respect, selective reflection spectroscopy which combines linearity and crossover-transitions-free sub-Doppler resolution could be useful, offering complementary information on transition amplitudes and positions for heavy atmospheric molecules.

In conclusion, we have performed high-resolution, linear spectroscopy of gas phase molecules at micrometric distances away from a surface. The achieved resolution is $\mathrm{\approx0.6\; MHz}$ limited by the laser linewidth but can be further improved by locking the QCL to a more stable frequency source \cite{Cappelli_OL, Sow_APL_2014, Argence_NatPhot_2015}. We demonstrate the advantages of this technique for enriching molecular databases and for molecule-surface interaction spectroscopy. 

This work, and its natural extension to molecular thin cells of sub-wavelength thickness paves the way towards:

1)The fabrication of simple and compact molecular frequency references throughout the spectrum without resorting to saturated absorption schemes, required in fiber platforms \cite{Triches2015, Knabe2009, Takiguchi2011}. Multiple cells \cite{NAUMOV2001207} or multipass techniques can increase the signal of such devices without compromising their compactness. 

2)The measurement of the molecule-surface interaction in nanometric thin cells that allow us to control molecular confinement by changing cell thickness \cite{fichet_epl_2007}. Probing molecule-surface interactions using rovibrational spectroscopy can be promising for measurement of the Casimir-Polder anisotropy \citep{Thiyam_PRA_2015}. This is because light induced transitions tend to orientate the molecule along the electric field of the probing beam \cite{Bimonte_PRA_2016}, while the electronic cloud remains in its ground state. Additionally, the interaction of molecules with near field thermal emission \cite{shchegrovprl2000, GreffetNature2002, laliotisnatcommun2014} can be a point of interest, as molecular rovibrational energy can be comparable to the thermal energy even at room temperatures. Finally, molecular electronic transitions can be used for exploring chirality effects \cite{Butcher_NJP_2012}. 

(3) Exploring the fundamental physics of subwavelength confinement with molecules. This includes studies of the Maxwell-Boltzmann distribution close to surfaces with narrow velocity selection \cite{todorov2017, Rabi_1994}, studies of supperradiance with molecules \cite{Feld_PRL_1973} or studies of local field corrections \cite{vuletic_optcomm_1993, guo_pra_1996, Peyrot_PRL_2018} with high density molecular gases. In this respect, the flexibility of molecular  cells that operate at room temperature with independent control of gas pressure, can be an additional advantage compared to atomic vapors.

\begin{acknowledgments}
We thank J. R. Rios Leite, M. Ducloy for discussions and A. Shelkovnikov, J. Grucker for participating in preliminary measurements. We acknowledge financial support from the ANR project SQUAT (Grant No. ANR-20-
CE92-0006-0.1), PVCM (Grant No.ANR-15-CE30-0005-01) and the Labex First-TF ANR 10 LABX 48 01.\\\\ \\\end{acknowledgments}
 
\bibliography{biblioMoleculesvBD}

\end{document}